\documentstyle[twocolumn,epsfig]{mn}
\newcommand{\mincir}{\raise
-2.truept\hbox{\rlap{\hbox{$\sim$}}\raise5.truept
\hbox{$<$}\ }}
\newcommand{\magcir}{\raise
-2.truept\hbox{\rlap{\hbox{$\sim$}}\raise5.truept
\hbox{$>$}\ }}
\newcommand{\minmag}{\raise-2.truept\hbox{\rlap{\hbox{$<$}}\raise
6.truept\hbox{$>$}\ }}
\newcommand{\be}{\begin{equation}}
\newcommand{\ee}{\end{equation}}
\newcommand{\ba}{\begin{eqnarray}}
\newcommand{\ea}{\end{eqnarray}}
\newcommand{\brr}{\begin{array}}

\newcommand{\err}{\end{array}}
\newcommand{\bc}{\begin{center}}
\newcommand{\ec}{\end{center}}

\title{The angular momentum of dark
halos: merger and accretion effects}
\author[S\'ebastien Peirani, Roya Mohayaee and Jos\'e A.\ de Freitas Pacheco]
{S\'ebastien Peirani, Roya Mohayaee
and Jos\'e A.\ de Freitas Pacheco
\\
Observatoire de la C\^ote d'Azur, B.P.4229, F-06304 Nice Cedex 4, France \\
emails: peirani@obs-nice.fr, roya@obs-nice.fr, pacheco@obs-nice.fr
}

\begin{document}

\maketitle

\begin{abstract}
We present new results on the angular momentum evolution of
dark matter halos. Halos, from N-body simulations, are classified 
according to their mass growth histories into two categories: 
the {\it accretion category} contains halos whose mass has varied 
continuously and smoothly, while the {\it merger category} contains
halos which have undergone sudden and
significant mass variations (greater than 1/3 of their initial mass per
event). We find
that the angular momentum  grows in both cases, well into the nonlinear regime.
For individual halos we observe strong correlation 
between the angular momentum variation and
the mass variation. The rate of growth of both mass and angular momentum
has a characteristic transition time at around $z \sim 1.5 - 1.8$, with
an early fast phase followed by a late slow phase. Halos of the 
merger catalog acquire more
angular momentum even when the scaling with mass is taken into account.
The spin parameter has a different behavior for the two
classes: there is a {\it decrease} with time for halos in the accretion catalog
whereas a small {\it increase} is observed for the merger catalog.
When the two catalogs are considered together, no significant variation of the
spin
parameter distribution with the redshift is obtained. We have also found that the spin
parameter
neither depends on the halo mass nor on the cosmological model.
From our simple model developed for the formation of a disk galaxy similar to
the Milky Way, we
conclude that our own halo must have captured satellites in order to
acquire the required angular momentum and to achieve most of the disk around
$z \sim 1.6$. The distribution of the angular momentum indicates that
at $z \sim 1.6$ only $22\%$ of the halos have angular momentum of magnitude comparable
to that of disk
galaxies in the mass range $10^{10}  - 5\times 10^{11} M_{\odot}$, clearly insufficient
to explain the present observed abundance of these objects.
\end{abstract}


\section{Introduction}

The origin of angular momentum remains a key factor in understanding the
formation,
evolution and particularly the morphological types of galaxies.
In the gravitational instability paradigm of structure formation, angular
momentum could arise from the tidal interaction of galaxies with their
surroundings (Hoyle 1949; Peebles 1969; Doroshkevish 1970; White 1984).
As a protogalaxy expands, its angular momentum grows linearly with time due to
tidal interactions until it decouples from the background expansion, turns
around and
starts collapsing. After the turn around time the angular momentum essentially
stops growing and becomes less sensitive to tidal torques (Peebles 1969),
and hence this time marks the maximum angular momentum acquired by galaxies
through this mechanism.

In a Universe dominated by dark matter, within the hierarchical scenario,
galaxies are formed  when baryonic gas falls into the gravitational potential
well
of  dark matter halos. Therefore,
many properties of galaxies, including their
angular momentum, are expected to be closely related to those of
their host halos. The acquisition of angular momentum by dark matter
halos through tidal torques, can be formulated within the Lagrangian
perturbation theory (White 1984; Catelan \& Theuns 1996a, 1996b). At the first
order,
using Zel'dovich approximation, one can show that the angular momentum
grows linearly with time.
This result has been confirmed by N-body simulations.
However, the tidal torque theory (TTT) is an oversimplified explanation
of the complex evolution of the angular momentum of dark halos.
Real progenitors of dark matter halos are not isolated, as assumed in TTT, but
are continuously growing by accretion and merger.
In the non-linear regime, no self-consistent theory, including  mass accretion
and merger history of halos yet exists. Most
of the investigations rely on numerical simulations and or on semi-analytic
modelings.
The transfer of orbital angular momentum to spin during a merger event
has been verified in simulations, where one observes a correlation between
a sharp increase in the halo mass and sudden variations in
the dimensionless spin
parameter $\lambda ={J\mid E \mid ^{1/2}}/{GM^{5/2}}$
where $J$ is the angular momentum, $E$ is the
total energy of the halo and $M$ is its mass.
The spin parameter is essentially the
ratio of the angular momentum of an object to that required for rotational
support. There also seems to be a general
tendency for $\lambda$ to decrease during periods of slow
accretion (Vitvitska et al. 2002). Effects of mergers are clearly seen in the
probability distribution of the spin parameter. Halos
that have undergone major merger events at $z \leq 2-3$ have systematically
larger spin
parameters than those that have evolved only by accretion (Gardner 2001; Vitvitska et
al. 2002;
Maller, Dekel \& Somerville 2002). Since on galactic scales ellipticals have
lower
angular momentum than spirals, these results could be troublesome if the former
are
merger remnants of the latter.

Thus, in spite of extensive works, many aspects of the evolution of
the angular momentum in the nonlinear regime remain inconclusive.
For instance, it remains unclear if the distribution of the spin  parameter
depends on the redshift or not (Lemson \& Kauffmann 1999;
Maller, Dekel  \& Somerville 2002). Concerning
the variation of $\lambda$ with the halo
mass, the early investigations indicate a
slight decrease with a mean slope
${d({\rm log}<\lambda>)}/{d({\rm log}M)} = -0.17\pm0.07$ (Barnes \& Efstathiou 1987).
Such a trend was also observed in more recent numerical simulations (Cole \&
Lacey 1996).
Although this result seems to be compatible with the tidal torque theory, a
small
but opposite variation was obtained if halos were to acquire angular momentum by merging
(Maller, Dekel \& Somerville 2002). Recent tests of TTT by N-body simulations
(Porciani, Dekel  \& Hoffman 2002a, 2002b) have shown that the values of the
spin
amplitudes predicted by TTT
are in agreement with those derived from simulations if the linear growth stops
somewhat earlier than maximum expansion. After shell crossing, some numerical
studies suggest that, in general, the angular momentum
decays (Barnes \& Efstathiou 1987; Sugerman, Summers \& Kamionkowski 2000),
which is expected due to spin conservation if the system remains isolated. On the other
hand, erratic variations are expected due to the transfer of orbital angular
momentum to
spin during merger events or due to accretion (Vitvitska et al. 2002).

In the present work, considering that models based on TTT do not agree in
detail with the results of N-body simulations in the nonlinear regime, we investigate an
alternative approach in which  halos obtain most of their spin through the
transfer of orbital angular momentum by continuous accretion and/or mergers.
We identify the halos using FOF algorithm (Davis et al. 1985) and 
make two distinct catalogs:
the ones which have never had a major merger event, and another catalog
for halos which have undergone at least one merger episode, in which their mass
increased by at least $1/3$ of their initial value during the event.
We show that even after shell crossing, the mean angular momentum
of dark halos still grows, but with a rate different from that predicted by TTT.
Individual
halos have erratic variations of the angular momentum, which are strongly
correlated with the mass accretion rate, producing on the average a growth
of the angular momentum. We recognize two scaling regimes for the growth of
the mean halo masses and angular momenta. The corresponding exponents differ
for the accretion and merger samples. We find that, when no distinction 
between the two catalogs
are made, i.e., when all the halos are considered together, no significant variation
of
the spin parameter distribution with the redshift is observed. However, when 
the halos are
considered separately, the evolution of spin parameter of the accretion sample
indicates a small but significant {\it decrease} with time, whereas for the merger
catalog the spin parameter {\it increases} with time. We show that this behavior is a
consequence of the different time scaling laws followed by the dynamical
variables defining the spin parameter.

Finally, we also discuss the formation of disk galaxies.
Past cosmological N-body/SPH simulations have shown that the scale length and
angular momenta of disks (baryonic component) are about 
one order of magnitude smaller than the
observed values. This problem is twofold:
firstly, in N-body simulations baryons loose a significant fraction of their 
angular momentum, resulting in disks that are too
small and secondly the angular momentum distributions
reveal too much low angular momentum material. If the specific angular
momentum distribution is conserved, which is usually assumed to be so for
disk formation, the resulting disk shows a more 
centrally concentrated mass distribution
making it difficult to explain
the exponential mass profile of spirals (Navarro, Frenk \& White 1995; Navarro
\& Steinmetz 1997, 2000; van
den Bosch 2001; van den Bosch, 
Burkert \& Swaters 2001). We discuss a simple
collapse model to form a baryonic proto-disk. We will show how the halo
parameters, in particular the specific angular momentum distribution, affect
the disk properties like its mass distribution, spin parameter and how the
angular momentum acquired during the growth phases of the halo influences the
evolution of the disk. The plan of this paper is as follows:
in Section 2, we review briefly the basic principles of TTT and alternative
theories,
in Section 3 we describe
our N-body simulations and our halo-finding algorithms and we
discuss the angular momentum evolution by accretion and merger
events in Section 4. In Section 5
we develop our scenario describing the formation of a disk galaxy and in
Section 6, we summarize our main results and conclusions.

\section{Angular Momentum: dynamical description}

The global angular momentum ${\bf J}$ of a halo of $N$ particles is defined by
\be
{\bf J}=\sum_{i=1}^N m_i {\bf r}_i \wedge {\bf v}_i\,,
\ee
where ${\bf r}_i$ and ${\bf v}_i$ are the position and velocity of the ith
particles with respect to the halo centre of mass.

In the continuum limit, the angular momentum ${\bf J}$ of the matter contained at
time $t$ in a volume $V$ of the Eulerian ${\bf x}$-space becomes
\ba
{\bf J}&=&\int_{a(t)^3V} \rho({\bf r}) d^3r\,\, ({\bf r}-\tilde{\bf r})\wedge v({\bf
r})\,, \nonumber\\
&=&
a(t)^5\int_V \rho({\bf x}) d^3x\,\, ({\bf x}-\tilde{\bf x})\wedge \dot x\,,
\nonumber\\
&=&
\bar \rho a(t)^5 \int_{V_q} d^3 q\,\, ({\bf q}+{\bf S} - \tilde{\bf q}
-\tilde{\bf S})\wedge
{d {\bf S}\over dt}\,,
\label{angmom}
\ea
where $a(t)$ is the scale factor, ${\bf x}$ is the comoving Eulerian spatial
coordinate, ${\bf r}=a(t){\bf x}$ is the physical distance, centre of mass
coordinates are marked by tilde, $\rho$ is the
matter density, $\bar\rho$ is the background density, $V_q$ is the
{\it Lagrangian} volume and ${\bf S}$ is the displacement vector from the
initial position ${\bf q}$ to the Eulerian position ${\bf x}$ satisfying the
mapping
\be
{\bf x}({\bf q},t)={\bf q}+{\bf S}({\bf q},t).
\ee
An analytic approximate
expression for the angular momentum (\ref{angmom}) can be found using
the linear Lagrangian perturbation theory, {\it i.e.} the Zel'dovich
approximation,
\be
{\bf S}({\bf q},t)\approx {\bf S}^{(1)}({\bf q},t)=D(t){\bf \nabla}\varphi({\bf
q})\,,
\label{s}
\ee
where $D(t)$ is the growing mode of the density perturbation: in the Einstein-de
Sitter Universe $D(t)\approx a(t)$ and $\varphi({\bf q})$ is the linear
gravitational potential satisfying to
first order in the density fluctuation, $\delta=(\rho-\bar\rho)/\bar\rho$ the
Poisson equation
$\nabla^2 \varphi=\delta/a$.
Inserting (\ref{s}) in the expression for the angular momentum
(\ref{angmom}), gives to first order
\be
J^{(1)}_i=\bar \rho a(t)^5 \dot D \int_{V_q} d^3q \epsilon_{ijk} (q_j-
\tilde q_j){\partial\varphi\over\partial q_k}\,.
\ee
Taylor expanding the gravitational potential around the centre of mass $\tilde
{\bf q}$, one obtains
\be
J^{1}_i\approx a(t)^5 \bar\rho \dot D \epsilon_{ijk} T_{kl} I_{lj}
\label{angmom1}
\ee
for the angular momentum in terms of the deformation tensor
\be
T_{kl}=\left( \partial^2\varphi\over\partial q_k\partial q_l\right)_{\tilde {\bf
q}}
\ee
and the quadrupole inertia tensor
\be
I_{lj}=\int_{V_q} d^3q (q_l-\tilde q_l)(q_j-\tilde q_j)\,.
\ee
Evidently, if the inertia and
deformation tensors are aligned then the tidal torque is zero. Numerical
simulations
indeed show that these two tensors are mostly aligned and the contribution to
the torque comes from statistical fluctuations in the misalignment (Porciani,
Dekel \& Hoffmann
2002a, 2002b). Thus, from expression (\ref{angmom1}), in the linear regime
angular momentum increases linearly with time. The prediction of the linear
theory were shown to be marginally affected by going to higher-order terms in
Lagrangian perturbation theory (Catelan \& Theuns 1996a).

If the Lagrangian volume is spherical, then the
moment of inertia is $I_{lj}=4\pi\delta_{lj} q^5/15$, where $\delta_{lj}$ is
the Kronecker delta, and hence the angular
momentum vanishes. This remains valid also at the second order in Lagrangian
perturbation theory (Catelan \& Theuns 1996b). However, the angular momentum
of an Eulerian spherical volume does indeed grow only at second order (Peebles
1969). The discrepancy between these two results was shown to be due to the
fact that the growth of the angular momentum of an Eulerian sphere at second
order is not due to tidal torques but due to the convective transport of the
angular momentum across the surface of the sphere (White 1984).

The subtlety in the generation of angular momentum comes from the fact that
in an expanding universe the vortical modes decay and using Kelvin circulation
theorem one can show that in a dissipationless system vorticity cannot be
generated, however, angular momentum can indeed be generated (Peebles
1973). Such motion which is mainly a shearing flow and not a vortical
flow, soon becomes complicated and vorticity can be generated. Furthermore,
a minute amount of velocity dispersion in the initial flow would indeed
lead to the generation of vorticity which would not show in cold dark
matter simulations where the velocity dispersion is put equal to zero.
A small vorticity can also
be obtained in
the averaged density-weighted streams after shell crossing (Pichon \& Bernardeau
1999). Moreover, a collisionless system with a non-isotropic velocity dispersion
can also generate vorticity (Ruzmaikin 1975).

Although TTT provides a good description of the growth of angular momentum in
the linear regime, as compared with numerical simulation, it is clearly an
oversimplified description of the dynamical process of spin acquisition.
The angular momentum of halos in this model is overestimated  by
a factor of three compared to simulations, with a large scatter. Also the
direction of the spin vector disagrees with numerical experiments, with errors
typically of about 57$^o$ (Lee \& Pen 2000).
TTT predictions for the spin parameter are for all the matter which at $z =0$
is found in a given halo and does not predict the spin of any particular halo
progenitor. Here, we follow the evolution of the angular momentum of the most massive
progenitor; a procedure more adequate for modeling galaxy
formation (Vitvitska et al. 2002, Somerville \& Primack 1999).

Alternative semi-analytic models for the acquisition of angular momentum by
halos exist; for example the model in which the spin is built up randomly 
by mass accretion and merger events (Vitvitska et al. 2002, 
Maller, Dekel \& Somerville 2002).
The evolution of the angular momentum in this scenario is rather different
from that expected from TTT, in which the halo spin grows linearly at
very early times and essentially stops later. The evolution of the
spin parameter of the major halo progenitor in such an alternative picture,
presents sharp increases during important  merger episodes and a tendency
to decline during periods of gradual accretion or capture of small lumps
of matter. The analytical approach of the random walk theory predicts
that the distribution of the spin parameter is a
log-normal, with mean and dispersion that depend
neither on the progenitor mass nor on the redshift.

\section{The simulations}

The N-body simulation analyzed uses adaptive particle-particle/particle-mesh
(AP$^3$M) code HYDRA (Couchman, Thomas \&
Pearce 1995).
We ran a $\Lambda$CDM  simulation
with $h=0.65$, $\Omega_m=0.3$, $\Omega_\Lambda=0.7$ and $\sigma_8=0.9$.
The simulation was performed in a periodic box of side $30h^{-1}$ Mpc
with $256^3$ particles (hence with mass resolution of $2.051 \times 10^8$
M$_\odot$).
The simulation started at $z=50$ and ended at the present time $z=0$.

Halo catalogs at different redshifts were prepared by using a
friends-of-friends
(FOF) algorithm (Davis et al. 1985), with a linking length of b = 0.15 in
units of the mean interparticle separation. Only
structures with at least $50$
particles were retained in the different catalogs, corresponding to a halo
mass of about
$1.03\times 10^{10} M_{\odot}$. Halos with masses higher than $10^{13}
M_{\odot}$
(about 49,000 particles) were excluded from our study, since they have
many substructures and are likely to be representative of groups or clusters of
galaxies.
Once a structure is identified, the total energy of each 
particle is computed, with respect to the centre of mass, and
those with positive energy are removed.
The procedure is repeated with the new  centre of mass, computed according
to its usual definition, until no unbound particles are found. We 
comment that our halo-finding
algorithm differs from those using the spherical virial radius around
the most bounded particle. Most of the halos characterized by this procedure
are not virialized, having a virial ratio ${2T}/{\mid W \mid}$ around
1.4 - 1.8, at $z \sim 3$. As we shall discuss in a forthcoming paper (Peirani et
al 2003), halos reach a state
of relaxation only by now, depending strongly on their evolutionary history.
The typical halo dimension was estimated calculating
the gravitational radius, $r_g$ = ${GM^2}/{\mid W\mid }$, where $W$ is the
total gravitational energy of the system.

We have identified $8728$ halos at $z=0$ and out of this number we have prepared
a sample of
$780$ halos which have never undergone a major merger event, and their mass have
always
increased by accretion or by capture of small lumps of matter.
We also selected $561$ halos which had at least one major merger episode
in their history, corresponding to an increase of their masses at least by a
factor of $1/3$ in the event. Since this limit, generally adopted in the
literature is rather arbitrary, we have also examined how our results are
modified if we decrease the above {\it mass-fraction threshold} to $1/6$. In this case, the
accretion catalog is reduced to 607 halos, whereas the merger sample is
increased to 734 halos. Here, we present our results mainly for the first set
of the catalogs, with mass-fraction threshold of $1/3$, however our
results are not significantly modified by the other set of catalogs with the
mass-fraction threshold of $1/6$. 

\section{Angular Momentum Evolution}

\subsection{Accretion effects}

In order to study  accretion effects,
we have selected halos which up to the present time have
not undergone any major merger event. This means that these halos have not
captured any other objects with masses larger than one third (one sixth) of their
own masses.
These halos are mostly in filaments and their masses evolve only by a continuous
accretion process or by the capture of small lumps of matter. The accreted
matter
comes from different directions and not only from the main axis, since filaments
are
much wider than the typical halo dimensions. The evolution of the selected halos
satisfying these conditions has been followed from $z = 3.5$ until $z = 0$,
using
the following algorithm: since each particle in the simulation can be
identified, it is
possible to obtain the constitution of each halo at each time step. If more than
70\%
of the particles of a given halo are found in another halo at the subsequent
time step, then
we assume that both objects are the same. This scheme allows the tracing of the
evolution
of the most massive progenitor. In order to follow the halo evolution,  the considered 
redshift interval ($3.5 \geq z \geq 0$) was divided  into 28 steps from which positions
and velocities of each particle were stored (specifically at:
0.0, 0.04, 0.09, 0.15, 0.20, 0.26, 0.33, 0.40, 0.48, 0.57, 0.67,
0.79, 0.92, 1.08, 1.26, 1.49, 1.6, 1.7, 1.77, 1.9, 2.0, 2.15, 2.3,
2.5, 2.68, 3.0, 3.2, 3.51).

The mass distribution of these $780$ halos (or the $607$ halos for the
catalogs with the mass-fraction
threshold of $1/6$) was calculated at different times and
statistical parameters like the mean and 
the median of the logarithm of mass, as well as
the root mean square deviation were computed. We emphasize that all of the  scaling 
relations obtained in this paper use the median of the logarithm of the
variable (which all vary over several decades) rather than the variable itself, 
since otherwise the statistical distribution would not have a fair representation.
Both the mean and the median increase with time, indicating
an average growth of about a factor of five in the considered time interval. As
we shall see later and as it is expected, merging produces on the average much
more massive halos.

In Fig. 1 we have plotted the 
variation of the median of the logarithm of the mass
of the halos, as a function of time. The simulated data is well-fitted by
the two power laws
\be
M_{acc} \propto\cases{
t^{0.91} \quad {\rm for} \quad  z > 1.8  \cr \cr
t^{0.58} \quad {\rm for} \quad  z < 1.8  \cr}
\label{massaccretion}
\ee
where $M_{acc}$ is the median of the halo mass 
distribution at a given instant of time.
If we consider the mass-fraction threshold of $1/6$ instead of $1/3$ to discriminate
between the two catalogs, these results are not significantly modified: the transition
between the two scaling regimes still occurs near $ z \sim 1.8$ and the
exponent at higher redshifts becomes 0.89, while for $z < 1.8$  the exponent
is 0.55.

\begin{figure}
\centerline{
        \vspace{-0.8cm}
        \epsfxsize=0.35\textwidth\rotatebox{-90}
        {\epsfbox{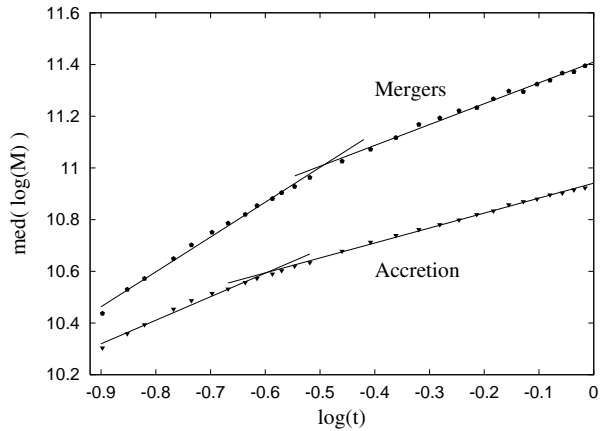}}
           }
\vspace{0.8cm}
\caption
{The evolution of the median of the logarithm of mass for the halos that
grow by merger and those which grow by accretion. Two scaling regimes are
observed. Time is given in unit of $H_0^{-1}$.}
\label{disc1}
\end{figure}

It is worth mentioning that this behavior compares with the evolution of
individual
halos (Vitvitska et al. 2002), indicating a fast growth of the mass
for $z > 1.5 - 2.0$, followed by a phase where the accretion rate is slower.
The median mass of the distribution fits the relation
$M_{acc}(z) =M_{acc}(0)e^{-{2z}/{(1+z_*)}}$,
where $M_{acc}(0)$ is the present halo mass
and $z_*$ is a characteristic redshift at which the halo mass is a certain
fraction of the present mass (Wechsler et al. 2002;
 Zhao et al. 2003). We found from our fit $z_* = 4.0$, indicating that at that
redshift the median mass was about five times smaller than now.

As we have seen in Section 2, tidal torques are effective until the
density perturbation attains the turnaround time, when they are expected
to reach the maximum angular momentum. However, for lumps of matter that
are continuously growing, the turnaround time is ill-defined since they
do not simply follow the evolution of the spherical model. In particular, in
the process of virialization, the inner mass shells cross the center and
one would expect an important energy transfer from bulk  to
random motions, due to collective effects. In fact, this mechanism
contributes to the relaxation of the halo towards a complete equilibrium
state, which is however delayed by the continuous accretion of matter.
Faced with this complex situation, some empirical
approaches have been suggested in the literature to define the moment of
the first shell crossing. Here, we take the time
of maximum velocity dispersion to define such an instant
(Sugerman, Summers \& Kamionkowski 2000).
It is worth mentioning that it was found
that, on the average, the magnitude of
the angular momentum, $J$, still grows almost linearly ($J
\propto t^{0.85}$)
between turnaround and the first shell-crossing time, as
defined above (Sugerman, Summers
\& Kamionkowski 2000).  We have estimated the latter for several
halos of different masses, $M$, and  obtained
that the first crossing time varies approximately with mass as
\be
t_c =0.112({M}/{10^{11}})^{0.41}
\ee
In this equation the time is in units of $H_0^{-1}$ and
masses are in solar units. We stress that here the halo masses correspond
to the instant of shell crossing.  At $z = 3.5$, when we start to follow the evolution of
the halos, the shell-crossing time of the median of the mass distribution
corresponds
to a redshift of about $z \sim 6$. Assuming roughly that the crossing time is
twice the turnaround time (this is correct only in the spherical model), the
latter
is around $z \sim 8-9$. In Fig. 2 we show
the estimated redshift of shell crossing by our procedure as a function of the halo mass.
This result implies that the variations
observed in the angular momentum of our simulated halos for $z < 3.5$ are probably
not due to the tidal field of nearby lumps of matter, but are due to
the transfer of angular momentum through accretion of mass.

\begin{figure}
\centerline{
        \vspace{-0.8cm}
        \epsfxsize=0.5\textwidth\rotatebox{0}
        {\epsfbox{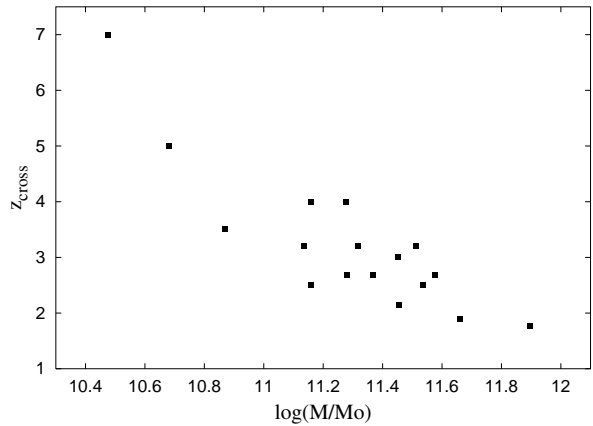}}
           }
\vspace{0.8cm}
\caption
{The redshift corresponding to the first shell-crossing time (defined as the
time of maximum velocity dispersion) as a function of the halo mass at that
redshift.}
\label{disc1}
\end{figure}

Variations of the angular momentum when halos are considered individually
are rather erratic and are correlated with variations of the mass accretion
rate.
This behavior is illustrated in Fig 3, where the rate of variation
of the modulus of the angular momentum is plotted in parallel with the mass
accretion rate for two halos from our catalogs. Notice that positive
and
negative rates can be observed
in both cases, but not always producing effects in the same sense, since they
depend on the relative orientation of the halo spin with respect to
the angular momentum vector of the accreted matter. This figure illustrates
that accretion affects considerably the angular momentum
history of the halos. It is worth mentioning that the FOF mechanism may
introduce some artifacts like bridges between nearby halos, which may
cause sudden variations in the mass accretion history. However our 
procedure always check if particles are physically 
bounded to the halo or not and in this way we reduce
the possibility that spurious effects would influence our results.

\begin{figure}
\centerline{
        \vspace{-0.8cm}
        \epsfxsize=0.5\textwidth\rotatebox{0}
        {\epsfbox{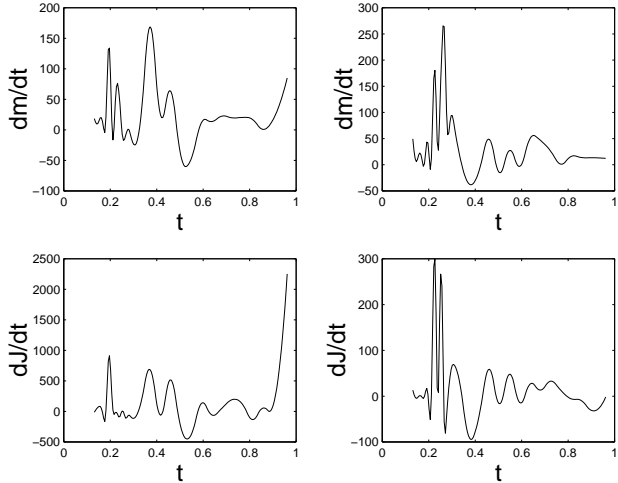}}
           }
\vspace{0.8cm}
\caption
{Two examples of the evolution of the angular momentum and mass
accretion rates. Time is given in units of $H_0^{-1}$. In both 
cases, there is a strong correlation between the two quantities. Example 1
corresponds to panels on the left while example 2 corresponds to panels
on the right.
}
\label{disc1}
\end{figure}

In spite of the erratic variations of the spin when individual halos are
considered,
on the average, an increase of the angular momentum is observed.
Fig. 4 shows the median value of angular momentum distribution,  med(${\rm log} J$),
as a function of time. Again, two scaling regimes are noticed with 
a characteristic redshift of $z \sim 1.8$.  A best fit to the data shows that
\be
J_{acc} \propto\cases{
t^{1.50} \quad {\rm for} \quad  z > 1.8  \cr \cr
t^{1.07} \quad {\rm for} \quad  z < 1.8  \cr}
\label{angularmomentumaccretion}
\ee
The results considering the mass-fraction threshold of $1/6$
are once again quite similar: the exponent
is $1.34$ for $z > 1.8$ and $0.97$ for $z < 1.8$.
From $z=3.5$ to $z=0$, halos gain, on average, one order
of magnitude in the mean angular momentum by such a mechanism.

\begin{figure}
\centerline{
        \vspace{-0.8cm}
        \epsfxsize=0.35\textwidth\rotatebox{-90}
        {\epsfbox{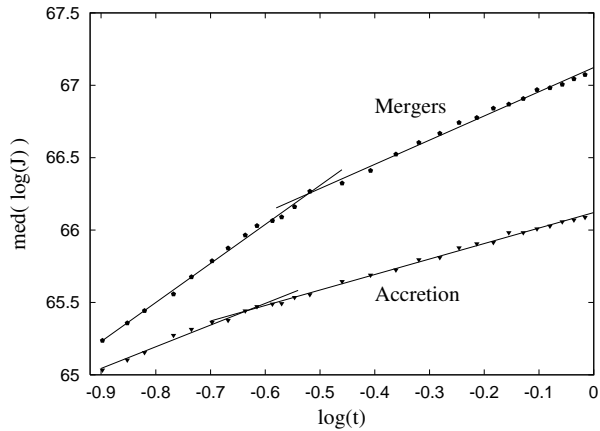}}
           }
\vspace{0.8cm}
\caption
{The evolution of the median of the logarithm of the 
angular momentum, ${\rm med} ({\rm log}\, J)$, with time
for the halos which grow by accretion and halos which have undergone at least
one merger. The growth continues well after the first shell-crossing time which
is at about $z\sim 6$ ($t\sim 0.065/H_0$).}
\label{disc1}
\end{figure}

\subsection{Merger effects}

In order to compare the effects of merging on the angular momentum history
of halos, we have selected 561 halos which have at least had one merger event
in which the mass-fraction threshold is greater than $1/3$. As before, to evaluate the
consequences of this threshold on the results, another merger catalog including 734
halos
was also studied, in which the threshold was lowered to $1/6$.

In Fig. 1 we show how the median of the logarithm of mass evolves for these halos. We
comment that
as compared to halos which grow by accretion, the median
grows faster, as expected, and we equally find two regimes and a
characteristic redshift of $z \sim 1.5$, namely

\be
M_{mer} \propto\cases{
t^{1. 35} \quad {\rm for} \quad  z > 1.5  \cr \cr
t^{0.81} \quad {\rm for} \quad  z < 1.5  \cr}
\label{massmerger}
\ee
Here again $M_{mer}$ corresponds to the median of the mass distribution at a
given redshift.
No significant modifications are noticed  if the mass-fraction threshold is taken equal
to $1/6$ in order to classify the two catalogs: the exponent is $1.25$ for $z > 1.5$
and decreases to $0.75$ for $z < 1.5$.

In Fig. 4 we have plotted the median (log J) of the angular momentum
distribution
as a function of time, which can be compared with the evolution of the halos
evolved only by accretion. Again, two regimes of growth are found with a
transition
at $z \sim 1.5$.

\be
J_{mer} \propto\cases{
t^{2.70} \quad {\rm for} \quad z > 1.5 \cr \cr
t^{1.67} \quad {\rm for} \quad z < 1.5 \cr}
\label{angularmomentummerger}
\ee

Notice the rapid rate of growth of the angular momentum due to merging for $z >
1.5$
and, in spite of increasing at a lower rate after the transition, the growth
is still faster than the almost linear variation obtained for the accretion
sample in the same evolutionary phase. No significant modifications are obtained
changing the mass-fraction threshold to $1/6$: the exponents become equal to 2.52 and
1.54 respectively before and after the characteristic redshift at $z \sim 1.5$. In the
redshift interval $3.5 > z \geq 0$, the average angular momentum grows by a
factor of $80$, suggesting clearly that merging transfer angular momentum much more
efficiently than accretion. However, a simple comparison as the one we have just made
is not sufficient to explain this phenomenon. In fact, since 
the angular momentum scales with
mass as $M^{5/3}$ (see below) and halos which have grown by merging are more
massive, this
effect must be taken into account when comparing both samples. In this case,
it is more convenient to
compare the distribution of the quantity ${\rm log} ({J}/{M^{5/3}})$. In Fig. 5
the distributions for the two samples at $z = 0$ are compared and one can
 see that merging events
give a larger contribution to the final angular momentum of halos irrespective
of the mass variations.

\begin{figure}
\centerline{
        \vspace{-0.8cm}
        \epsfxsize=0.5\textwidth
        {\epsfbox{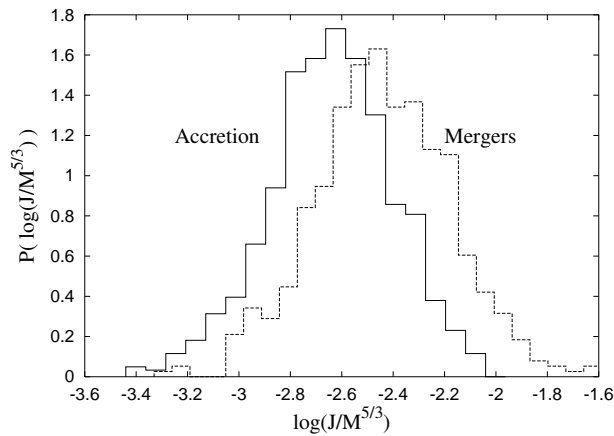}}
           }
\vspace{0.8cm}
\caption
{The distribution of the angular momentum, normalized by $M^{5/3}$, for halos
that
have grown by merging (mass ratio $1/3$) and those which never had a major
merger
event. The mean of the distribution is larger for halos which 
have undergone major merger events.}
\label{disc1}
\end{figure}

In the context of the present study, an interesting question, which we have
also considered was the radial redistribution of the specific angular momentum during
accretion or after a merger episode. In the accretion process, the
specific angular momentum is distributed almost linearly with the distance to the centre
($j_{DM} \propto r$), and the amplitude at the periphery may increase or
decrease with time, according to the relative orientations of the halo spin and that
of the matter stream being captured. In the upper panel of Fig. 6, the radial
distribution of the specific angular momentum is shown at three different 
redshifts, for a typical case where there is a small decrease of the
maximum value, in spite of the increasing halo dimensions. However, notice that 
the distribution becomes more and more
smooth, which is a general trend of the evolution when accretion dominates. The situation is
rather different after an important merger event, and this is illustrated
in the lower panel of Fig.6 . After the merger episode, the maximum
specific angular momentum increases, which is a quite general situation.
In the subsequent phases, the specific angular momentum is redistributed,
decreasing in the central regions and increasing in the halo outskirts. 
We shall return to this problem in Section 5, when the formation
of a disk galaxy will be discussed. However, to characterize  well these
effects, higher resolution simulations are needed.

\begin{figure}
 \begin{center}
    \rotatebox{0}{ \includegraphics[height=6cm,width=8cm]{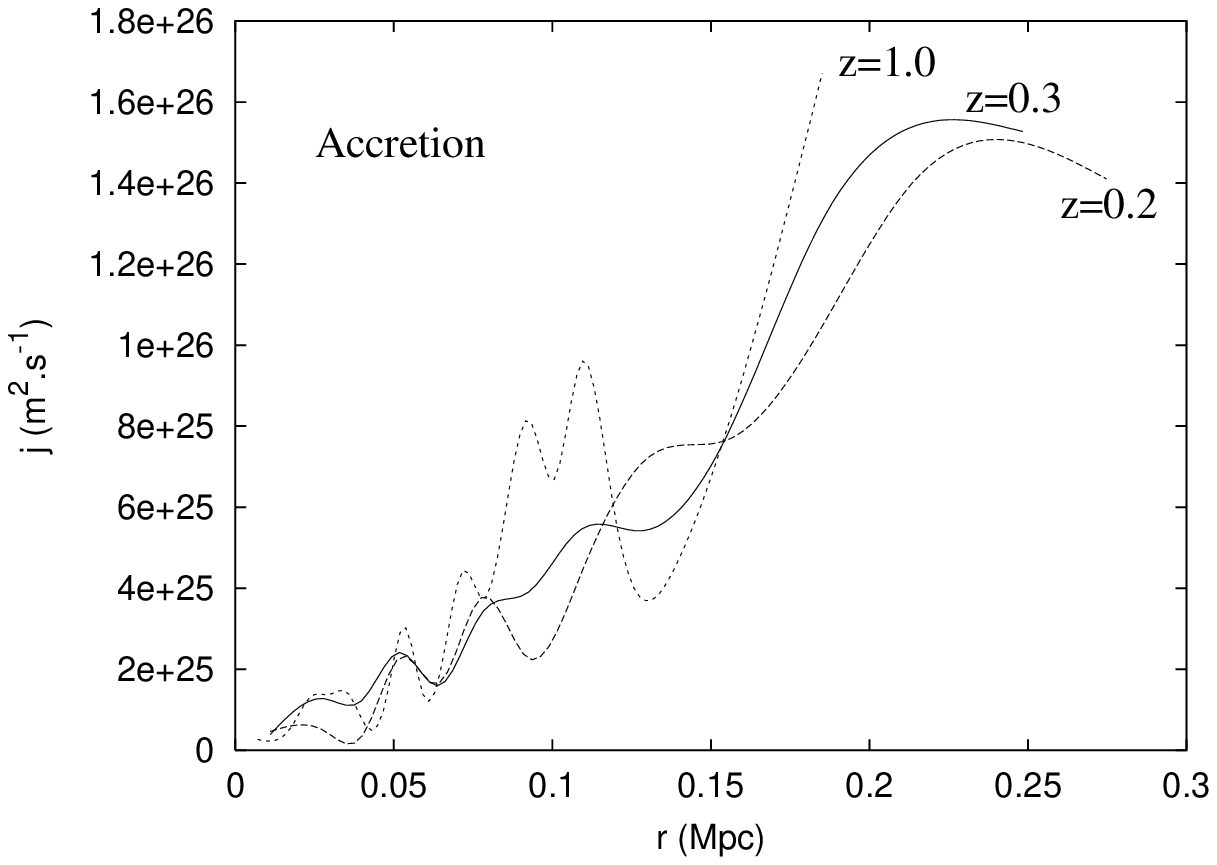}}
  \end{center}
  \vfill
  \vspace{-8mm}
  \begin{center}
    \rotatebox{0}{ \includegraphics[height=6cm,width=8cm]{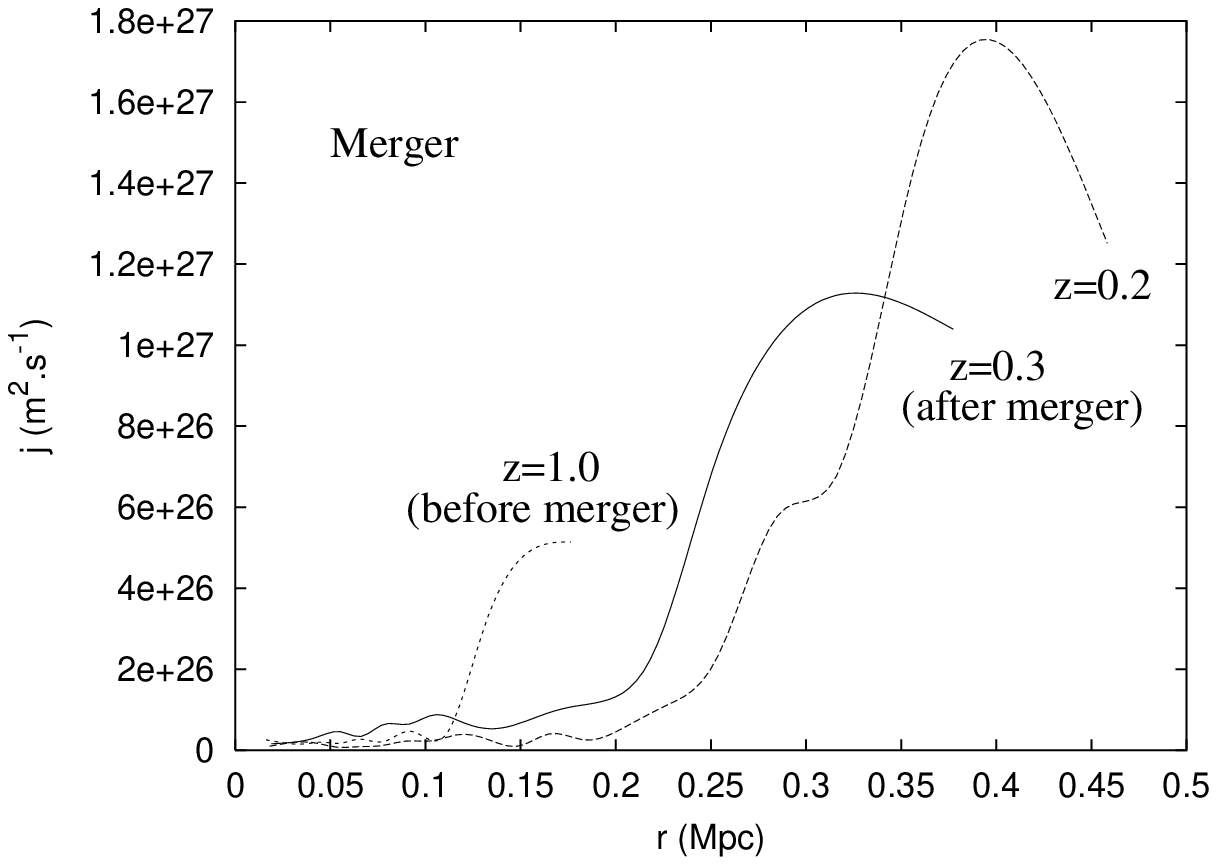}}
  \end{center}
\caption
{The distribution of the specific angular momentum for a typical halo of $12000$
  particles at $z=0$ in the accretion case (upper panel) and of $30,000$ particles 
at $z=0$  in the merger case (lower panel).  In the merger case, there is a 
general trend of transfer of angular momentum from the central regions
to the periphery.}
\label{figure6}
\end{figure}

\subsection{The evolution of the spin parameter}

\begin{figure}
\centerline{
        \vspace{-0.8cm}
        \epsfxsize=0.5\textwidth
        {\epsfbox{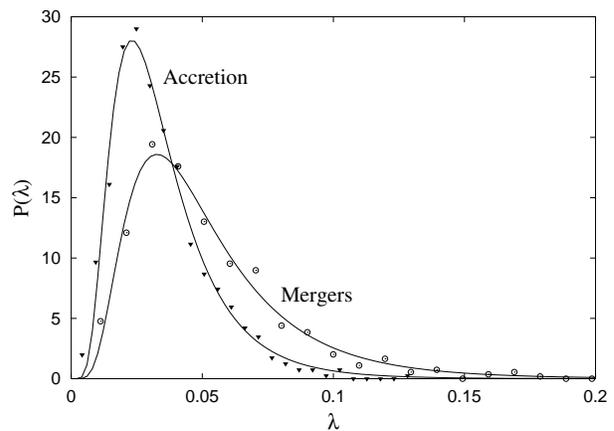}}
           }
\vspace{0.8cm}
\caption
{The distribution of the spin parameter for halos that grow by merger and those
which grow by accretion at $z = 0$. The average spin parameter is higher for
halos underwent
merger events than for halos which have growth by accretion only.}
\label{fig6}
\end{figure}

Previous works have shown that the spin parameter $\lambda$, obtained from
simulations, has a log-normal distribution (Barnes \& Efstathiou 1987; Cole
\& Lacey 1996; van den Bosch 1998; Ryden 1998),

\begin{equation}
P(\lambda)d\lambda =
\frac{1}{\sigma_{\lambda}\sqrt{2\pi}}exp\left(-\frac{ln^2(\lambda/
\lambda_0)}{2\sigma^2_{\lambda}}\right)\frac{d\lambda}{\lambda}
\end{equation}
which seems to be a universal result, independent of the
cosmological model. In Fig. \ref{fig6} we show the distribution of the spin parameter at
$z = 0$
for the two catalogs considered here. An inspection of
this plot confirms
again that halos which have undergone important merger
episodes have, on the average, a larger spin parameter and a wider distribution
 than those evolved by accretion only.
We will return to this point later in this section.

If the halos of both samples are considered together, the parameters defining
the distribution at $z =0$ are: $\lambda_0 = 0.036$ and $\sigma_{\lambda} =
0.57$, in
agreement with previous studies (Warren et al. 1992; Gardner 2001; Bullock
et al. 2001).

\begin{figure}
  
  \begin{center}
    \rotatebox{0}{ \includegraphics[height=6cm,width=8cm]{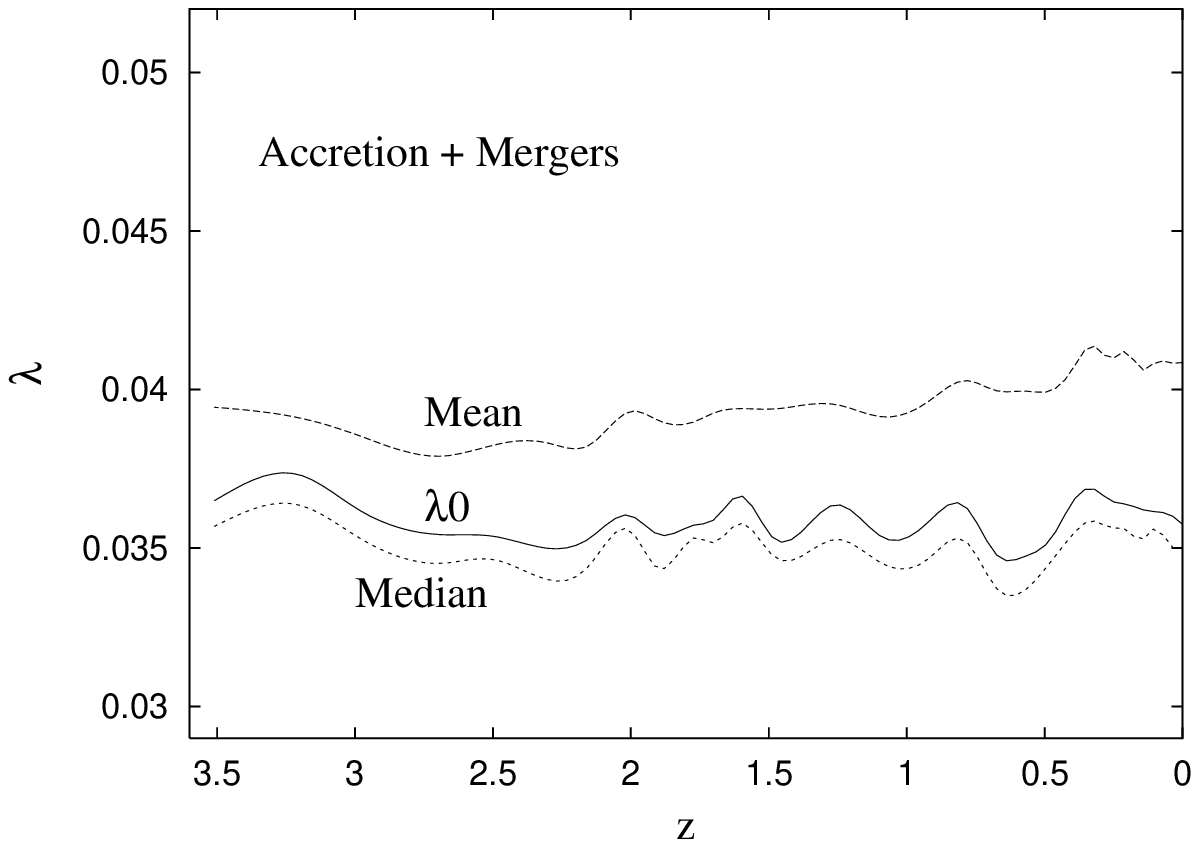}}
  \end{center}
  \vfill
  \vspace{-8mm}
  \begin{center}
    \rotatebox{0}{ \includegraphics[height=6cm,width=8cm]{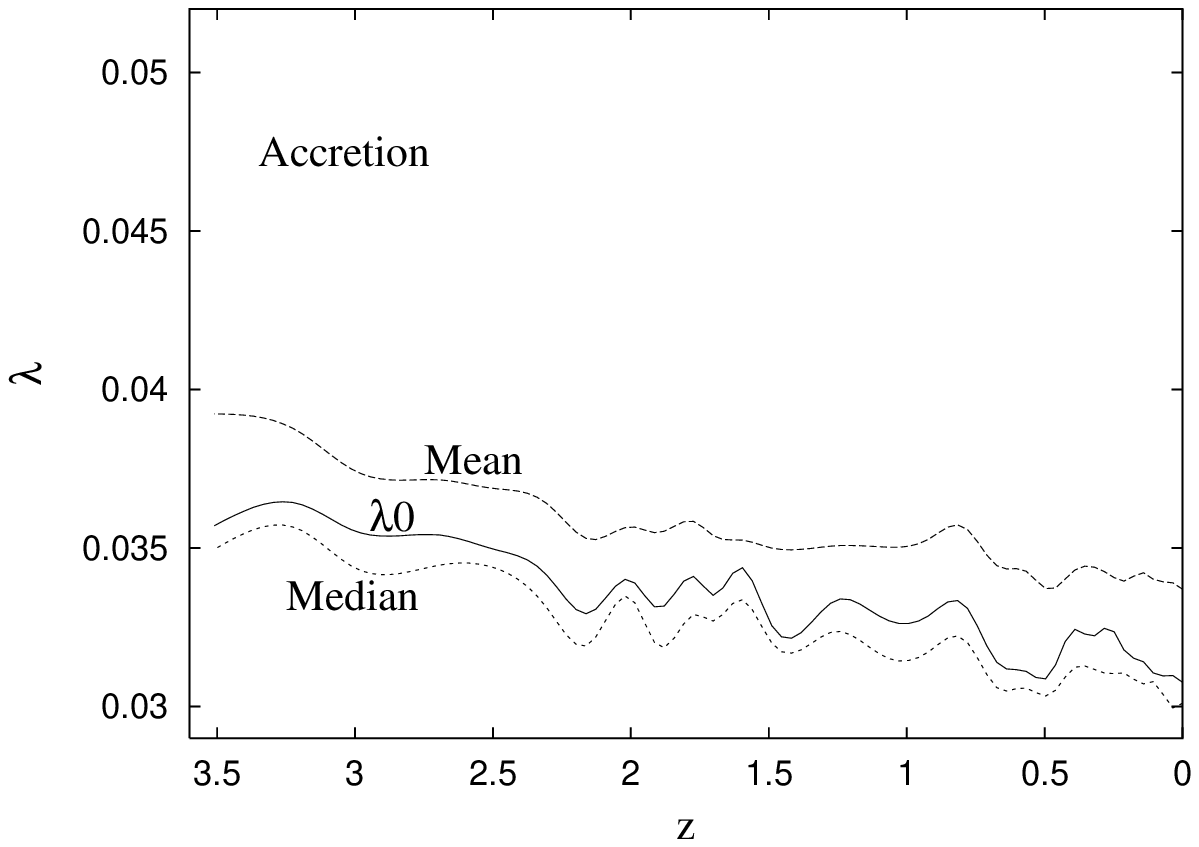}}
  \end{center}
  \vfill
  \vspace{-8mm}
  \begin{center}
    \rotatebox{0}{ \includegraphics[height=6cm,width=8cm]{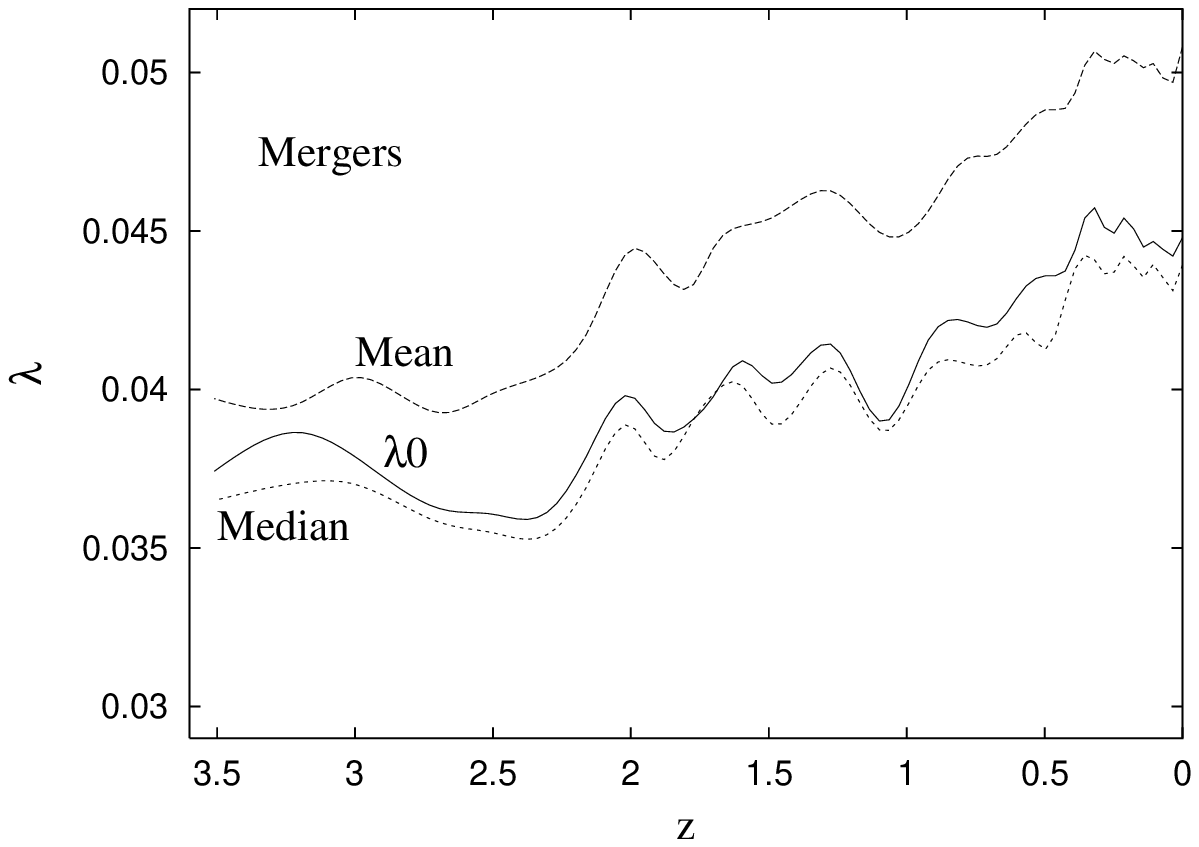}}
  \end{center}

\caption
{Panels from up  to down: a) evolution of the spin parameter
for all the halos (accretion + merger). Three statistical
parameters are shown: the median, the mean and
$\lambda_0$; evolution of the same parameters but for halos of the accreting
catalog only (b) and for halos of the merger catalog (c).}
\label{fig8}
\end{figure}

Earlier investigations of the evolution of the spin parameter 
indicate that its distribution does not vary
with redshift. This result seems to be valid both in the TTT and in the
random walk capture model (Vitvitska et al. 2002). In Fig. \ref{fig8} we plot statistical
parameters of the $\lambda$ distribution, as the mean, median and $\lambda_0$,
as a function of the redshift, when the halos of each catalog separately and also both
catalogs together are considered ($1341$ halos).

This figure demonstrates that, in spite of some erratic fluctuations,
the variations in both $\lambda_0$ and the median are quite small,
although
a slight increase seems to be suggested by the evolution of the mean value.
Taken
at the face value, these results are compatible with no variation of $\lambda$
with
$z$ or, at least compatible with a slight increase in the mean. However, the
results are quite different if we analyze the evolution of the spin parameter
for the
two samples (accretion and merger) independently. The middle and the  lower
panels of Fig. \ref{fig8} show respectively
the evolution of the same statistical parameters for halos evolved only by
accretion
and those which have had at least one major merger event. We notice
a clear and significant evolution in opposite senses:  halos evolved by
accretion
have a {\it decreasing} spin parameter, whereas for halos evolved
by merging $\lambda$ {\it increases}. Typical values for the rmsd of the mean
value of the spin parameter $\lambda$ are 0.02.

\begin{figure}
\centerline{
        \vspace{-0.8cm}
        \epsfxsize=0.35\textwidth\rotatebox{-90}
        {\epsfbox{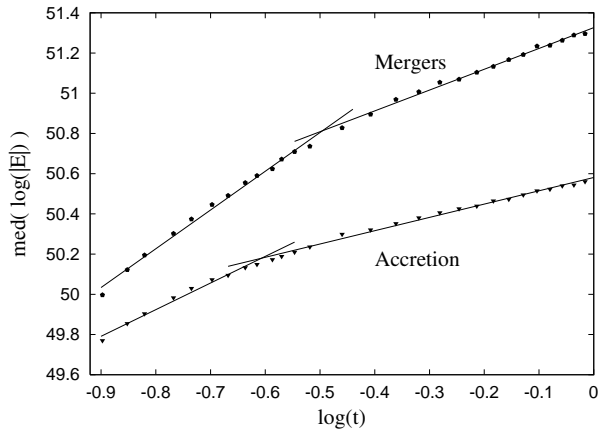}}
         }
\vspace{0.8cm}
\caption
{The evolution of the median of the logarithm of the modulus of the total energy
$\mid E\mid$ of the halos separately from the accretion and merger catalogs}
\label{fig9}
\end{figure}

How can this behavior be explained ? The
spin parameter
depends on dynamical quantities since $\lambda \propto { J |E|^{1/2}}/
{M^{5/2}}$. As we
have seen above, both the angular momentum $J$ and the mass $M$
increase with
time at different rates, according to the dominant mechanisms by which halos
grow.
But what about the total energy. In Fig. \ref{fig9} we have plotted the
median of the logarithm of the modulus of the total
energy as a function of time. We notice that halos evolved by merging have
a higher binding energy that also evolves faster than in the accretion case.

This question
will be discussed in more detail by Peirani et al. (2003), when the 
dynamical relaxation of the halos will be examined.
For the evolution of the magnitude of the energy, once again we observe
two scaling regimes. For the accretion catalog, we have
\be
|E|_{acc} \propto\cases{
t^{1.32} \quad {\rm for} \quad z > 1.8 \cr \cr
t^{0.66} \quad {\rm for} \quad z < 1.8 \cr}
\ee
and for the merger catalog we have
\be
|E|_{mer} \propto\cases{
t^{1.93} \quad {\rm for} \quad z > 1.5 \cr \cr
t^{1.03} \quad {\rm for} \quad z < 1.5 \cr}
\ee
Next, we consider the evolution for $z < 1.8$, and put the above results together
with the scaling relations in
(\ref{massaccretion}),(\ref{angularmomentumaccretion}),
(\ref{massmerger}) and (\ref{angularmomentummerger}).
In the accretion case we have $J
\propto t^{1.07}$,
$M \propto t^{0.58}$ and $|E| \propto t^{0.66}$, which imply $\lambda
\propto t^{-0.05}$. The same scaling analysis for $z > 1.8$ gives $\lambda \propto
t^{-0.12}$.
Therefore the scaling regimes of the different dynamical variables lead to a
decrease of
the spin parameter. In the case of merging, for $z < 1.5$, $J \propto t^{1.67}$,
$M \propto t^{0.81}$ and
$|E|\propto t^{1.03}$, implying $\lambda \propto t^{0.16}$, or an
increase of
the spin parameter with time (for $z > 1.5$ we have $\lambda \propto t^{0.29}$),
in agreement
with the results of our simulations.
Thus, in order to explain the evolution of the spin parameter, the scaling
regimes
of the three dynamical quantities  $J, M$ and $E$ are needed to be considered.

A similar analysis may be performed in the case of the variation of the spin
parameter with the halo mass. The total energy scales with mass as
$|E| \propto M^{\kappa}$ and the angular momentum as $J \propto
M^{\gamma}$. From a theoretical point of view, both exponents should  be
equal to {\it 5/3}, and consequently the spin parameter should not depend on
the mass. From our simulations, we have calculated these exponents and
have verified that: i) they practically do not vary with the redshift and are quite
close to the expected theoretical  value; ii) no variation is also detected
when the results of both samples are compared, since for  halos
evolved by accretion the mean values are $\kappa = 1.647 \pm 0.006$ and
$\gamma = 1.647 \pm 0.034$, while for halos having experienced merger
events the mean values are $\kappa = 1.624 \pm 0.011$ and $\gamma =
1.672 \pm 0.023$. These results imply $\lambda \propto M^{-0.02}$, consistent
with a very small dependence on the halo mass. Using simulations with lower
mass resolution ($128^3$ particles), we have checked if the scale parameters
$\kappa$ and $\gamma$ depend on the initial conditions, as claimed by
Barnes \& Efstathiou (1987) and no differences where noticed either using
a $\Lambda$CDM or a CDM model.

\section{Formation of disk galaxies}

Mestel (1963) proposed that the distribution of matter in a collapsed
disk galaxy could be determined, if each element of the proto-galaxy
conserves its specific angular momentum. Previous, N-body/SPH simulations
have showed that the scale length and specific angular momentum of
simulated disks are one order of magnitude smaller than observed, the
so-called "angular momentum catastrophe" (Navarro \& Benz 1991; Navarro
\& White 1994). More recent simulations by van den Bosch et al. (2002) seem to
confirm that both gas and dark matter have similar angular momentum
distributions, but
their spin vectors are, in general, misaligned.

In this section, in spite of not being the major goal of this work, we explore
the
formation of disk galaxies in connection with
the results of our simulations on the angular momentum of halos. The bulge
formation
will not be considered here, since it may be
considered as a separate system, once its collapse timescale is probably
as short as that of ellipticals of comparable mass. Moreover, they host in
general
a supermassive black hole whose  r\^ole in the bulge formation is not yet
well established (see, for instance, Silk 2001). From a dynamical
point of view, it has been suggested that if the globular cluster
system and the bulge of our Galaxy have similar dynamics, then the bulge
would have a specific angular momentum about a third that of the
disk (Frenk \& White 1980).

In the hierarchical scenario, galaxies are expected to be formed inside
dark matter halos. As discussed above, halos may initially acquire angular
momentum  by tidal torques
until they reach the turnaround  phase. Then the tidal field decreases
and the angular momentum increases as a consequence of accretion and merging
episodes. In spite of the fact that our simulations include only dark matter, we
would expect that mechanisms affecting the intrinsic angular momentum of
such a component will also act on  baryons. In this case, one would expect
that the specific angular momentum of the gas (baryons) is proportional
to that acquired by the dark matter, namely, $j_b = \beta j_{dm}$, where
$\beta \sim 1$ is a parameter describing the efficiency of torques on the gas
component.

Here a simple picture is developed, based on our simulations, to describe the
resulting baryonic disk structure. We have chosen two examples to illustrate
a possible evolutionary history of our own Galaxy. Both halos have masses
at $z = 0$ comparable to that of our Galaxy ,
but have evolved differently. The first one has grown without  having any
major
merger episode, while the second one captured some small satellites (with
masses less than $1/3$ of the mass of the main halo)  between
$2.0 > z > 1.0$. In Table 1  we give the main evolutionary characteristics of
both examples.

The total angular momentum of the galactic (baryonic) disk is estimated to
be $J_b \approx 2.0 \times 10^{67}  kg.m^2.s^{-1}$ (see, for instance, Fall
\& Efstathiou 1980). As we will see below, the baryonic to dark matter
mass ratio in our Galaxy is taken to be $f_b \approx 0.034$ (
which is obtained from tuning
the model to roughly match the rotation curve of Milky Way). Thus if both
components have the same specific angular momentum, the required total
angular momentum of the halo must be (at least) $J_{dm} \approx 5.8 \times 10^{68}
kg.m^2.s^{-1}$. Simple inspection of Table 1 shows that the first example
attains
the necessary angular momentum only by the present time, unable to form a disk
10-11 Gyr old.  The second example, thanks to the capture of small halos, the
required angular momentum is attained at $z \sim 1.6$, corresponding to an age
of 10.4 Gyr, according to our adopted cosmological parameters. In what follows 
we consider the history of example 2 as 
representative of  the  halo in which the Milky Way in embedded.

The build up of the disk is gradual, following the growth of the halo
mass and the increase of the angular momentum. If we suppose arbitrarily
that the gas starts to settle
down towards the equatorial plane, according to its specific angular
momentum, at $z = 5$,  the halo has a mass of about $3.6 \times 10^{11}
M_{\odot}$ and a
gravitational radius of only 39 kpc! The halo is not yet completely relaxed,
since its
virial ratio is about 1.42.
\vspace{0.5cm}
\hspace{-0.6cm}{\underline{halo 1: growth by accretion only}}
\begin{flushleft}
\begin{tabular}{|c|c|c|c|}
z & $r_g  ({\rm kpc}) $ & ${\rm log} (M/M_{\odot})$ & ${\rm log} J ({\rm
  kgm}^2/{\rm s})$ \\
5.00 & 35.5  & 11.371  &\hspace*{-1cm} 66.017 \\
1.77 & 112.4 & 12.130  &\hspace*{-1cm} 67.780 \\
0.92 & 144.2 & 12.285 &\hspace*{-1cm} 67.905 \\
0.48 & 183.0 & 12.354 &\hspace*{-1cm} 68.253 \\
0.00 & 247.6 & 12.492 &\hspace*{-1cm} 68.778 \\
\end{tabular}
\end{flushleft}
\vspace{0.7cm}
{\underline{halo 2: growth by merger in $2 > z > 1$}}
\begin{flushleft}
\begin{tabular}{|c|c|c|c|}
z &  $r_g ({\rm kpc})$ & ${\rm log} (M/M_{\odot})$ & ${\rm log} J ({\rm
  kgm}^2/{\rm s})$ \\
5.00 & 39.0  & 11.556 &\hspace*{-1cm} 66.720 \\
1.77 & 137.7 & 12.324 &\hspace*{-1cm} 68.389 \\
0.92 & 197.3 & 12.510 &\hspace*{-1cm} 68.971 \\
0.48 & 226.3 & 12.550 &\hspace*{-1cm} 68.569 \\
0.00 & 278.1 & 12.628 &\hspace*{-1cm} 68.867 \\
\end{tabular}
\end{flushleft}
{\small Table 1: Possible halo progenitors of the Galaxy.}
\vspace{0.5cm}

Instead of studying the radial distribution of the specific angular momentum,
we consider the distribution in concentric cylindrical shells at a distance $l$
from
the spin axis $\vec J$. This approach is more adequate to study an axi-symmetric
disk, and can be justified by the results of Bullock et al. (2001), who have
concluded
that the distribution of the angular momentum in cells
tends to be more cylindrical than spherical, with a weak dependence on the
co-latitude $\theta$. Moreover, this has the advantage  that the integral
\begin{equation}
J_{dm} = \int^{R_{dm}}_0 2\pi j_{dm}(l)\Sigma_{dm}(l)ldl
\end{equation}
gives the modulus $J_{dm}$ of the total angular momentum of the halo (dark matter).
In
this equation, $\Sigma_{dm}(l)$  is the column density at a distance $l$ from the
spin
axis and $R_{dm}$ is the halo radius.

\begin{figure}
\centerline{
        \vspace{-0.8cm}
        \epsfxsize=0.5\textwidth\rotatebox{0}
        {\epsfbox{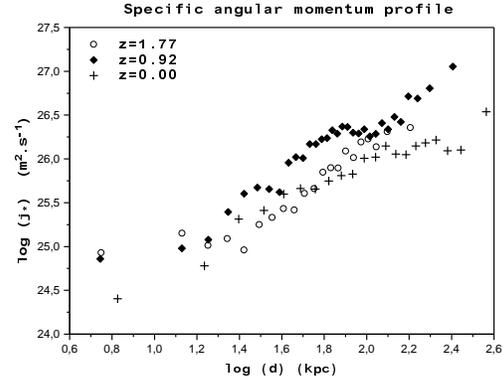}}
        }
        \vspace{1.1cm}
\caption
{The specific angular momentum $j_{dm}$ of dark matter calculated in concentric
cylindrical shells as a function of the distance to the spin axis and at
different redshifts.}
\label{fig10}
\end{figure}

In Fig. \ref{fig10} we show the profile of the specific angular momentum of dark matter,
which
can be quite well fitted by a power law $j_{dm}(l) = kl^{\alpha}$. The exponent
$\alpha$
varies with the redshift,  but typically is in the interval 0.9 - 1.5. These profiles
compare quite well with those obtained by Bullock et al. (2001), who tested
the cylindrical symmetry by similar plots.

The gas is initially supposed to be distributed as the
dark matter. It is worth mentioning
that if this is the case and both components  have initially the same specific
angular
momentum ($\beta$ = 1), then the ratio between the spin parameter of dark matter
$\lambda_{dm}$ to that of the baryons $\lambda_b$, is approximately
$\lambda_{dm}/ \lambda_b \approx f_b$. Notice that, in this case, the
initial spin parameter of the gas can be almost one order of  magnitude higher
than that of dark matter and not equal.
However, it should be emphasized that this conclusion depends on the 
adopted definition of the spin parameter. If one defines $\lambda^{'} =
j/\sqrt{2}R_{vir}V_{vir}$, where $R_{vir}$ and $V_{vir}$ are respectively
the virial radius and the circular rotation velocity at that point (see, for
instance, Bullock et al. 2001),  the ratio of $\lambda^{'}_{dm}/\lambda^{'}_b$ 
will be equal to the unity if both components have the same specific angular
momentum. Here we keep the original definition (Peebles 1969).

Since the gas cools, the quasi-equilibrium situation cannot be maintained
and the disk will be built up from the successively infalling shells. If a
dynamical equilibrium
situation is reached after the collapse, the dimension of the disk is fixed by
the maximum value of the specific angular momentum. Supposing $\beta$ = 1, the
rotational equilibrium can be expressed as
\begin{equation}
j_{dm}(l) =  j_b(y) = y\sqrt{\frac{GM_{dm}(y)}{y} + y\frac{\partial
\phi_b(y)}{\partial y}}
\end{equation}
where $y$ is the distance to the spin axis in the disk {\it after the collapse},
$M_{dm}(y)$ is the dark matter mass inside a radius y, $\phi_b(y)$ is the
gravitational
potential of  baryons, supposed to be distributed in a thin disk and given by
the
equation
\be
\phi_b(y) = - 2\pi G \int^{\infty}_0 dk J_0(ky)\int^{\infty}_0
dx\Sigma_b(x)J_0(kx)x
\ee
where $J_n(x)$ is the  Bessel function of order n and $\Sigma_b(y)$ is the
projected
baryon mass density of the disk.

\begin{figure}
\centerline{
        \vspace{-0.8cm}
        \epsfxsize=0.5\textwidth\rotatebox{0}
        {\epsfbox{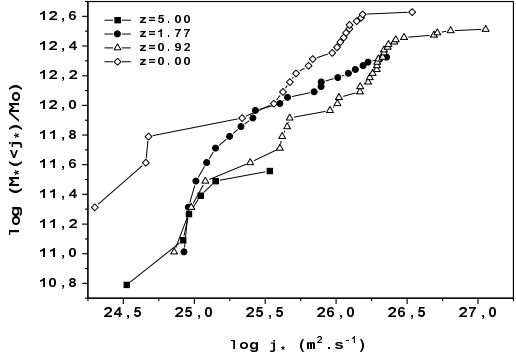}}
        }
         \vspace{1.1cm}
\caption
{The distribution of the projected dark matter mass with a maximum specific angular momentum
$j_{dm}$.}
\label{fig11}
\end{figure}

In order to derived the resulting density distribution $\Sigma_b(y)$, we have
adopted
the following procedure:

a) from our simulations, we compute not only the specific angular momentum
profile
but also the total dark matter mass inside a given radius r, the total projected
mass inside
a distance l from the spin axis or, equivalently the total projected mass with
specific
angular momentum less than $j_{dm}$. The latter distribution at  different
redshifts
is showed in Fig. \ref{fig11} and, generically, is represented by $M_{dm, p} = g(j_{dm})$,
where the extra subscript {\it p} means the projected mass.

b) Using the condition of dynamical equilibrium, the baryon mass in the disk
inside a distance y from the axis is
\be
M_{b,p}(y) =  f_b g(j_b(y))
\ee
and the derivative of this equation gives the projected  baryon
density
\be
\Sigma_b(y) = \frac{1}{2\pi y}(\frac{dM_{b,p}(y)}{dy})
\ee

c) since in the beginning the baryon distribution throughout the disk
is not known, we start our
calculations supposing that the gravitational field is essentially due to the
dark matter. Then, an initial projected density is calculated and the
corresponding gravitational potential (eq. 19), which is introduced in the
dynamical
equilibrium equation. The process is repeated until the required convergence
precision is attained.

\begin{figure}
\centerline{
        \vspace{-0.8cm}
        \epsfxsize=0.5\textwidth\rotatebox{0}
        {\epsfbox{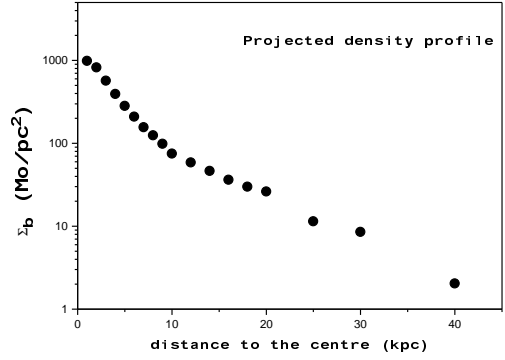}}
        }
        \vspace{1.0cm}
       \caption
{The resulted mass density profile of the baryonic disk.}
\label{fig12}
\end{figure}

\begin{figure}
\centerline{
        \vspace{-0.8cm}
        \epsfxsize=0.5\textwidth\rotatebox{0}
        {\epsfbox{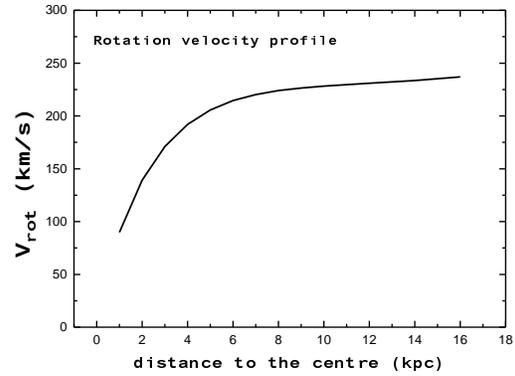}}
        }
        \vspace{1.0cm}
\caption
{Theoretical rotational velocity profile, including the contribution of baryonic
and dark matter.}
\label{fig13}
\end{figure}

If we start at redshift $z = 5$, a quite well-known result is obtained. The halo
does not have sufficient 
angular momentum and the resulting disk is quite small,
with
a radius of about 5 kpc. As the halo grows, its angular momentum increases and
the outer layers acquire a higher specific angular momentum. If the gas follows
the
same trend a larger disk can be built  up. Fig.\ref{fig12} shows the resulting projected
density profile. The projected mass density decreases exponentially with a scale
length
of 3.4 kpc between 1 - 10 kpc. Beyond 12 kpc the density decreases more slowly
but still exponentially, with a scale length of
about 8.4 kpc. The present extension of the disk
is approximately 40 kpc, where a density cutoff is obtained and its total mass
(baryonic) is about $1.45 \times 10^{11}
M_{\odot}$.  The rotation curve imposes a strong constraint on the
baryon fraction, which  should be equal to $f_b$ = 0.034, in order that the
rotation
velocity be about 230 km/s at a distance of 8-9 kpc from the spin axis (see Fig.
\ref{fig13}).

The required baryon fraction is a factor 4.5 lower than the mean cosmic value,
but
not incompatible with the ratio estimated for the Galaxy and the value
adopted by Bullock et al. (2001) in their own calculations of the disk
structure. Recent hydrodynamical
calculations by van den Bosch, Abel \& Hernquist (2003) considered the effects
of preheating on galaxy formation. In particular, a fraction of the preheated
gas may
become unbounded, yielding a final baryon to dark matter ratio smaller than
the mean cosmic value.

As we have mentioned above, if baryons and dark matter have the same specific
angular momentum and distribution, we would expect that the spin parameter
ratio between both components is $\lambda_{dm}/\lambda_b \approx f_b$. Once
that the disk structure was calculated, the total energy can be computed. For
an exponential disk, $E = -5.8G\Sigma_0^2\Lambda^3$, where $\Sigma_0$ is
the central projected mass density and $\Lambda$ is the scale of length. Using
the values derived from our model, we obtain $\mid E \mid = 5.5 \times 10^{51}$
J
and a spin parameter $\lambda_b = 0.49$. In this case, taking into account the
derived  baryon mass fraction, the expected halo spin parameter is
$\lambda_{dm} \approx 0.017$. It is worth mentioning that the value derived from
our simulations, when the halo (and the disk) reached about 90\% of its final
mass is $\lambda_{dm} = 0.018$, in agreement with our prediction.

\section{Conclusions}

We have studied the effects of accretion and merger on the
dynamical
evolution of dark matter halos, specifically on 
their angular momentum. The difference between accretion and merger
could
essentially be semantical, since the criterion for separating the
two cases is somehow arbitrary. In this work, accretion should  be understood as a
"continuous
and almost smooth variation" of the halo mass, while this variation in the
merging case is  "sudden and significant". When a halo mass changes by
more than a $1/3$ of its present value then we consider the event a merger and
otherwise accretion. In this manner, we have built and studied two separate
merger and accretion catalogs.  We have demonstrated that our results 
are only marginally modified if the
mass-fraction threshold of $1/3$ is reduced by a factor of two.

We found that in both cases (accretion and merger), there are two distinct
phases of mass
growth: an early and a fast phase, followed by a late phase of slow mass growth,
({\it e.g}.\ in agreement with Zhao et al. 2003).  The transition 
occurs at $z \sim 1.5-1.8$ (later than the value of $z \sim 3$ found
by Zhao et al. 2003). The  mass growth in both phases are 
well-represented by power laws. The exponents not only vary from one phase to
another, but
also change between the merger and the accretion catalogs. It is worth 
mentioning that in the hierarchical picture, if the power spectrum
of the initial fluctuations is of the form $\mid \delta_k \mid ^2 \propto k^n$,
in the linear regime the characteristic masses grow as $m \propto t^{4/(n+3)}$
(Peebles 1980). In this case, for a Harrison-Zel'dovich spectrum, the masses
at the early phases grow almost linearly with time, an evolution close to that
found from our simulations for halos in the merger catalog. A similar exponent
was also found by Toth \& Ostriker (1992), i.e., $M \propto t^{1.3}$, who
considered the halo mass evolution in the  hierarchical picture, using a standard
 CDM power spectrum.

The algorithm used in this work is adequate to follow the evolution of the more
massive
halo progenitor either in the accretion or merger case. Therefore, our
procedure is more
suitable to test the evolution of the angular momentum, for example 
according to the random walk
capture scenario (Vitvitska et al. 2002). It should be emphasized
that unlike the present work, in that model no distinction between accretion 
and merger is made and essentially the
effects of a cumulative capture of satellites by the progenitor is considered.

When considered individually,  halos show  erratic variations of the angular
momentum
strongly correlated (positively or negatively) with mass variations (Figs. 3).
This correlation reinforces the fact that even after 
the first shell-crossing the angular
momentum
varies, and the reason is mainly the transfer of orbital angular momentum to the
halo
spin due to accretion and merger. Previously, a different evolution for
individual halos in the simulations was found: in general the angular momentum decays
after the first shell crossing (Sugerman, Summers \& Kamionkowski 2000). It
was then understood that such a trend is probably due to the
redistribution of the angular momentum to bound particles outside the
overdensity
cut-off generated in the identification of the objects included in their
catalog.
Thus, unless all bound particles of a halo are identified, the apparent decay of
the angular momentum at late times can arise as a numerical artifact.

The angular momentum distribution calculated at different redshifts indicate
that
the median and the mean value increase with time. Two regimes again are
identified
in both samples, with a transition at $z \sim 1.5-1.8$, consistent with the
behavior observed
in the mass growth.  The two regimes of growth are again well represented by
power
laws (eq. 11) and are quite different from the linear variation expected from
TTT.
The present mass weighted distribution of the angular momentum
(Fig. 5) and the distribution of the spin parameter (Fig.\ref{fig6}) indicate that halos
which have undergone merger episodes gain more angular momentum, confirming some
previous investigations (Gardner 2001).

The distinction  between the two cases (accretion and merger) allows us to
establish an
important characteristic of the temporal behavior of the spin parameter. The
median, the mean and $\lambda_0$ increase with time for halos which have undergone major
merger episodes, while a decrease is obtained for the accretion catalog.
We have shown that such a behavior can be understood taking into account
the different scaling laws followed by the dynamical variables defining the spin
parameter. When no distinction is made, i.e., all halos are considered together,
the statistical parameters of the $\lambda$ distribution are practically
independent
of the redshift, apart from the mean which shows a slight increase with time.
This result is
consistent with studies in which no separation concerning the halo growth
history is made (see, for instance, Lemson \& Kauffmann 1999).

Concerning the possible dependence of the spin parameter on the halo mass
(Barnes \& Efstathiou 1987), we found that for $z < 3.5$ the angular momentum
and total energy scale with mass approximately as $M^{5/3}$, i.e., the
expected theoretical exponent. This results is valid both in $\Lambda$CDM and
in CDM cosmologies.  In this case, the spin parameter should be
practically independent of the halo mass and of its previous growth history,
since the same result was found for both catalogs.

We have also developed a simple model for disk 
formation based on considerations of the 
specific angular momentum distribution. 
The basic assumptions are the equality  between the specific
angular momentum of dark matter and baryons, and rotational equilibrium after
the collapse. However,  the distribution of the dark matter mass with the specific
angular momentum  was calculated here in a cylindrical symmetry, centered along
the spin axis. Two possible  progenitors of our own galactic halo were chosen
based
on their present mass. The first example concerns a halo evolved under accretion
only, which is
unable to acquire enough angular momentum to explain the observed value of the
galactic disk. In the second example, the halo was taken from the merger
catalog where it has captured a few satellites in
the
redshift interval $2 > z > 1$ and the required angular momentum is reached at
$z \sim 1.6$. The disk is built up gradually, the inner 5 kpc is formed around
 $z \leq 5$, and the outskirts attain
dimensions of about 30 kpc and 60\% of the total mass at $z \sim 1.6$, corresponding
to an age of about 10.4 Gyr. Only 17\% of the total mass was accreted in
the last 5 Gyr. This evolutionary path is probably not unique, since other
halos with different histories can also form disks
with similar characteristics. However, the general
picture agrees with the work by Helmi, White \& Springel (2003), who also
consider that the galactic  halo was formed gradually,  with more than
60\% of the mass already present at around $11$ Gyr ago. The resulting projected
mass density is exponential within the first 10 kpc, having a scale of length
equal to 3.4 kpc (Fig.\ref{fig12}).  The spin parameter of the disk is
$\lambda_b \sim 0.49$, in agreement with the relation
$\lambda_{dm}/\lambda_b \approx f_b$, expected if the specific angular momenta
of baryons and dark  matter are equal.

Finally, to conclude, it is important to notice that most disk galaxies  have
masses
in the range $10^{10} - 5 \times 10^{11} M_{\odot}$. Considering that the
angular
momentum of baryonic and dark matter scales as $M^{5/3}$, this translates into the
fact
that the halos associated to galaxies in that mass interval should be in the
range
$3 \times 10^{66}$ up to $2 \times 10^{69}  kg.m^2.s^{-1}$. If disks have ages
comparable to our own galactic disk, i.e., have been formed around $z \sim 1.6$,
our simulations indicate that only 22\% of the associated halos have acquired,
at that redshift,
the required angular momentum. If one neglects the eventual loss of spirals
that have merged to form ellipticals, the estimated fraction is
about a
factor of three less than the present observed value. This is perhaps the
''true'' angular momentum problem !

\vspace{1.0cm}

\noindent
{\bf Acknowledgement}

\noindent
We thank the referee for the useful comments which have contributed to improve
the text of this paper.
S.\ P.\ acknowledges PhD fellowship from Universit\'e de Nice Sophia-Antipolis
(UNSA). R.\ M.\ is supported by Marie
curie fellowship HPMF-CT 2002-01532.



\begin{thebibliography}{}

\bibitem[87]{be}
Barnes J.\ \& Efstathiou G.\ , 1987, ApJ 319, 575
\bibitem[2001]{bullock}
Bullock J.\ S.\ , Dekel A.\ , Kolatt T.\ S.\, Kravtsov A.\ V.\ , Klypin A.\
A.\ , Porciani C.\ \& Primack J.\ R.\ , 2001, ApJ 555, 240
\bibitem[1996]{cta}
Catelan P.\ \& Theuns T.\ ,1996a, MNRAS 282, 436
\bibitem[1996]{ctb}
Catelan P.\ \& Theuns T.\ ,1996b, MNRAS 282, 455
\bibitem[1996]{Co}
Cole S.\ \& Lacey C.\ , 1996, MNRAS 281, 716
\bibitem[1995]{}
Couchman H.\ M.\ P.\ , Thomas P.\ A.\ \& Pearce F.\ R.\ 1995, ApJ 452, 797
\bibitem[1985]{}
Davis M.\ , Efstathiou G.\ , Frenk C.\ S.\ \& White S.\ D.\ M.\ , 1985, ApJ 292, 371
\bibitem[1970]{d}
Doroshkevich A.\ , 1970, Astrofizika 6, 581
\bibitem[1980]{}
Fall S.\ M.\ \& Efstathiou G.\ , 1980, MNRAS 193, 189
\bibitem[1980]{}
Frenk C.\ S.\ \& White S.\ D.\ M.\ , 1980, MNRAS 193, 295
\bibitem[2001]{gardner}
Gardner J.\ P.\ , 2001, ApJ 557, 616
\bibitem[2003]{}
Helmi A.\ , White S.\ D.\ M.\ \& Springel V., MNRAS 339, 834
\bibitem[1949]{Ho}
Hoyle F., 1949, in Problems of Cosmological Aerodynamics, eds. Burgers J.M., van
der Hulst H.C., Central Air Documents Office, Dayton OH
\bibitem[2000]{}
Lee J.\ H.\ \& Pen U.-L.\ , 2000, ApJ 532, L5
\bibitem[1999]{lk}
Lemson G.\ \& Kauffmann G.\ , 1999, MNRAS 302, 111
\bibitem[2002]{MA}
Maller A.\ H.\ , Dekel A.\ \& Somerville R.\ , 2002, MNRAS 329, 423
\bibitem[1963]{}
Mestel L.\ , 1963, MNRAS 126, 553
\bibitem[1991]{}
Navarro J.\ F.\ \& Benz W.\ , 1991, ApJ 380, 320
\bibitem[1994]{}
Navarro J.\ F.\ \& White S.\ D.\ M.\ , 1994, MNRAS 267, 401
\bibitem[1997]{}
Navarro J.\ , Frenk C.\ S.\ \& White S.\ D.\ M.\ , 1995, MNRAS 275, 720
\bibitem[1997]{}
Navarro J.\ F.\ \& Steinmetz M.\ , 1997, ApJ 478, 13
\bibitem[2000]{}
Navarro J.\ F.\ \& Steinmetz M.\ , 2000, ApJ 538, 477
\bibitem[1969]{p}
Peebles P.\ J.\ E.\ , 1969, ApJ 155, 393
\bibitem[1973]{p}
Peebles P.\ J.\ E.\ , 1973, PASJ 25, 291
\bibitem[1980]{}
Peebles P.\ J.\ E., 1980 in Large-Scale Structure of the Universe, Princeton Series
in Physics, Princeton University Press
\bibitem[2003]{}
Peirani, S.\ , Mohayaee R.\ \& de Freitas Pacheco J.A., 2003, in preparation
\bibitem[1999]{}
Pichon C.\ \& Bernardeau F.\ , 1999, A\&A 343, 663
\bibitem[2002]{}
Porciani C.\ , Dekel A.\ \& Hoffmann Y.\ , 2002a, MNRAS 332, 325
\bibitem[2002]{pdhb}
Porciani C.\ , Dekel A.\ \& Hoffmann Y.\ , 2002b, MNRAS 332, 339
\bibitem[1975]{}
Ruzmaikin A.\ A.\ , 1975, Sov.Ast.Lett. 1, 95
\bibitem[1988]{ryden}
Ryden B.\ S.\ , 1988, ApJ 329, 589
\bibitem[2001]{Silk}
Silk J.\ ,  2001, astro-ph/0109325
\bibitem[1999]{}
Somerville R.\ S.\ \& Primack J.\ R.\ , 1999, MNRAS 310, 1087
\bibitem[2000]{Su}
Sugerman B.\ , Summers F.\ J.\ \& Kamionkowski M.\ , 2000, MNRAS 311, 762
\bibitem[1992]{}
Toth G. \& Ostriker J.P. , 1992, ApJ 389, 5
\bibitem[1998]{}
van den Bosch F.\ C.\ , 1998, ApJ 507, 601
\bibitem[2001]{vandenbosch}
van den Bosch F.\ C.\ , 2001, MNRAS 327, 1334
\bibitem[2001]{}
van den Bosh F.\ C.\ , Burkert A.\ \& Swaters R.A., 2001, MNRAS 326, 1205
\bibitem[2002]{}
van den Bosch F.\ C.\ , Abel T.\ , Croft R.\ A.\ C.\ , Hernquist L.\ \& White
S.\ D.\ M.\ , 2002, ApJ 576, 21
\bibitem[2003]{}
van den Bosch F.\ C.\ , Abel T.\ \& Hernquist L.\ , 2003, astro-ph/0308117
\bibitem[2002]{vitvitska}
Vitvitska M.\ , Klypin A.\ , Kravtsov A.\ V.\ , Wechsler R.\ H.\ , Primack J.\
R.\& Bullock J.\ S.\ , 2002, ApJ 581, 799
\bibitem[1992]{}
Warren M.\ S.\ , Quinn P.\ J.\ , Salmon J.\ K.\ \& Zurek W.\ H.\ , 1992, ApJ 399, 405
\bibitem[2002]{}
Wechsler R.\ H.\ , Bullock J.\ S.\ , Primack J.\ R.\ , Kravtsov A.\ V.\ \&
Dekel A.\ , 2002, ApJ 568, 52
\bibitem[1984]{w}
White S.\ D.\ M.\ , 1984, ApJ 286, 38
\bibitem[2003]{zhao}
Zhao D.\ H.\ , Mo H.\ J.\ , Ping Y.\ P.\ \& B\"orner G.\ , 2003, MNRAS 339,
12


\end{thebibliography}
\end{document}